\documentclass[aps,pre,preprint,groupedaddress]{revtex4}
\usepackage[dvips]{graphics}
\begin{document}

%\preprint{}

\title{\bf Symmetry pattern transition in cellular automata \\ with complex behavior}


\author{\bf Juan R. S\'{a}nchez,}
%\email[]{jsanchez@fi.mdp.edu.ar}
%\thanks{}
\affiliation{Facultad de Ingenier\'{\i}a - Universidad Nacional de Mar del Plata,\\
7600 Mar del Plata - Argentina.}

\author{\bf Ricardo L\'opez-Ruiz}
\affiliation{DIIS and BIFI, Facultad de Ciencias, Universidad de Zaragoza, \\
50009 Zaragoza, Espa\~{n}a.}

%%\date{}

\begin{abstract}
A transition from asymmetric to symmetric patterns in time-dependent extended systems is  described.
It is found that one dimensional cellular automata, started from fully random initial conditions, 
can be forced to evolve into complex {\it symmetrical} patterns by stochastically 
coupling a proportion $p$ of pairs of sites located 
at equal distance from the center of the lattice. 
A nontrivial critical value of $p$ must be surpassed in order to obtain symmetrical patterns 
during the evolution. This strategy is able to classify the cellular automata rules 
-with complex behavior- between those that support time-dependent symmetric patterns 
and those which do not support such kind of patterns.
\end{abstract}

% \pacs{05.45.1b, 05.40.1j, 05.70.Fh, 64.60.2i}

%\keywords{}

\maketitle


The stability analysis of patterns in extended systems has been revealed 
to be a difficult task. The many nonlinearly interacting degrees of freedom can 
destabilize the system by adding small perturbations to some of them. The impossibility 
to control all those degrees of freedom finally drives the dynamics toward 
a complex spatio-temporal evolution. Hence, it is of a great interest to develop
techniques able to compel the dynamics toward a particular kind of structure. 
The application of such techniques forces the system to
approach the stable manifold of the required pattern, and
then the dynamics finally decays to that target pattern. 

Synchronization strategies in extended systems can be useful in order to achieve such goal.
Different types of synchronization have been described in the literature.
They have been applied with success to synchronize lattices of iterated maps, 
coupled ordinary differential equations and cellular automata (CA). 
In this note, the stochastic synchronization method introduced in 
reference \cite{Morelli1998} for two CA is specifically used to find symmetrical patterns
in the evolution of a single automaton. To achieve this goal a stochastic operator, below described,
is applied to sites symmetrically located from the center of the lattice. 
It is shown that a {\it symmetry} transition take place in the spatio-temporal pattern. 
The transition forces the automaton to evolve toward complex patterns that 
have mirror symmetry respect to the central axe of the pattern. In consequence,
this synchronization method can be interpreted as a control technique for 
stabilizing complex symmetrical patterns.      

Cellular automata are extended systems, discrete both in space and time. 
The simplest type of automaton is a one-dimensional string composed 
of $N$ sites or cells. Each site is labeled by an index $i=1,\ldots,N$, 
with a local variable $s_i$ carrying a binary value, either $0$ or $1$.  
The set of sites values at time $t$ represents a configuration 
(state or pattern) $\sigma_t$ of the automaton.  During the automaton evolution, 
a new configuration $\sigma_{t+1}$ at time $t + 1$ is obtained by the 
application of a rule or operator $\Phi$ to the present configuration:
\begin{equation} \label{eq1}
\sigma_{t+1} = \Phi\:[\sigma_{t}]\:.
\end{equation}  
Usually, a local coupling among the nearest neighbors is implemented.
The state of site $i$ at time $t+1$, $s^{t+1}_i$, is a function of the value 
of the site itself at time $t$ and the values of its neighbors $s_{i-1}^{t}$ 
and $s_{i+1}^{t}$ at the same time. Thus, the evolution equation (\ref{eq1}) 
can be locally expressed as $s_i^{t+1} = \phi(s_{i-1}^t,s_i^t,s_{i+1}^t)$, 
being $\phi$ the particular realization to the site level of the rule $\Phi$.
As there are $2^3$ possible different local configurations as inputs, 
this means that there exist $2^8$ different possible evolution rules. 
Each one of these rules generates a different dynamical two-dimensional spatio-temporal pattern.
Wolfram already classified the different structures obtained from the 256 rules 
in four great groups ~\cite{Wolfram1983}. The interested reader is addressed to the original 
reference where this development can be followed.  

Our present interest resides in those CA evolving under rules capable
to show asymptotic complex behavior (rules of class III and IV). 
The technique applied here is similar to the synchronization scheme introduced by Morelli 
and Zanette~\cite{Morelli1998} for two CA evolving under the same rule $\Phi$. 
The strategy supposes that the two systems have a {\it partial} knowledge one about 
each the other. At each time step and after the application of the rule $\Phi$,
both systems compare their present configurations $\Phi[\sigma^1_t]$ 
and $\Phi[\sigma^2_t]$ along all their extension 
and they synchronize a percentage $p$ of the total of their different sites.  
The location of  the percentage $p$ of sites that are going to be put equal is 
decided at random and, for this reason, it is said to be an stochastic synchronization.
If we call this stochastic operator $\Gamma_p$,
its action over the couple $(\Phi[\sigma^1_t],\Phi[\sigma^2_t])$ can be represented 
by the expression:
\begin{equation} \label{eq2}
(\sigma^1_{t+1},\sigma^2_{t+1}) = \Gamma_p(\Phi[\sigma^1_{t}],\Phi[\sigma^2_{t}])=
(\Gamma_p\circ\Phi)(\sigma^1_{t},\sigma^2_{t}).
\end{equation}  

The same strategy can be applied to a single automaton with a even number of sites. 
Now the evolution equation, $\sigma_{t+1} = (\Gamma_p\circ\Phi)[\sigma_t]$, 
given by the successive action of the two operators $\Phi$ and $\Gamma_p$, 
can be applied to the configuration $\sigma_t$ as follows:  
\begin{enumerate}
	\item the deterministic operator $\Phi$ for the evolution of the automaton
	produces $\Phi[\sigma_t]$, and,
	\item the stochastic operator $\Gamma_p$, produces the result 
	$\Gamma_p(\Phi[\sigma_t])$, in such way that, if sites symmetrically located 
	from the center are different, i.e. $s_{i} \neq s_{N-i+1}$, then 
	$\Gamma_p$ equals $s_{N-i+1}$ to $s_i$ with probability $p$. 
	$\Gamma_p$ leaves the sites unchanged with probability $1-p$.  
\end{enumerate}

A simple way to visualize the transition to a symmetric pattern can be 
done by splitting the automaton in two subsystems ($\sigma^1_t,\sigma^2_t$), 
\begin{itemize}
	\item $\sigma^1_t$, composed by the set of sites $s(i)$ with $i=1,\ldots,N/2$ and
	\item $\sigma^2_t$, composed the set of symmetrically located sites $s(N-i+1)$ with $i=1,\ldots,N/2$, 
\end{itemize}
and displaying the evolution of the difference automaton (DA), defined as
\begin{equation}
\delta^{t}=\mid\sigma_t^1 - \sigma_t^2 \mid \:.
\end{equation}
The mean density of active sites for the difference automaton, defined as
\begin{equation}
\rho^t={2\over N}\sum_{i=1}^{N/2}\delta^t
\end{equation}
represents the Hamming distance between the sets $\sigma^1$ and $\sigma^2$. 
It is clear that the automaton will display a symmetric pattern when 
$\lim_{t\rightarrow\infty}\rho^t=0$. 
For class III and IV rules, a symmetry transition controlled by the parameter $p$ is found.
The transition is characterized by the DA behavior:
\begin{eqnarray*}
\hbox{when $p<p_c$} & \rightarrow & \hbox{$\:\lim_{t\to\infty}\rho^t \neq 0$ (complex non-symmetric patterns),} \\
\hbox{when $p>p_c$} & \rightarrow & \hbox{$\:\lim_{t\to\infty}\rho^t=0$ (complex symmetric patterns)}.
\end{eqnarray*}
The critical value of the parameter $p_c$ signals the transition point.  

\begin{figure}
	\includegraphics{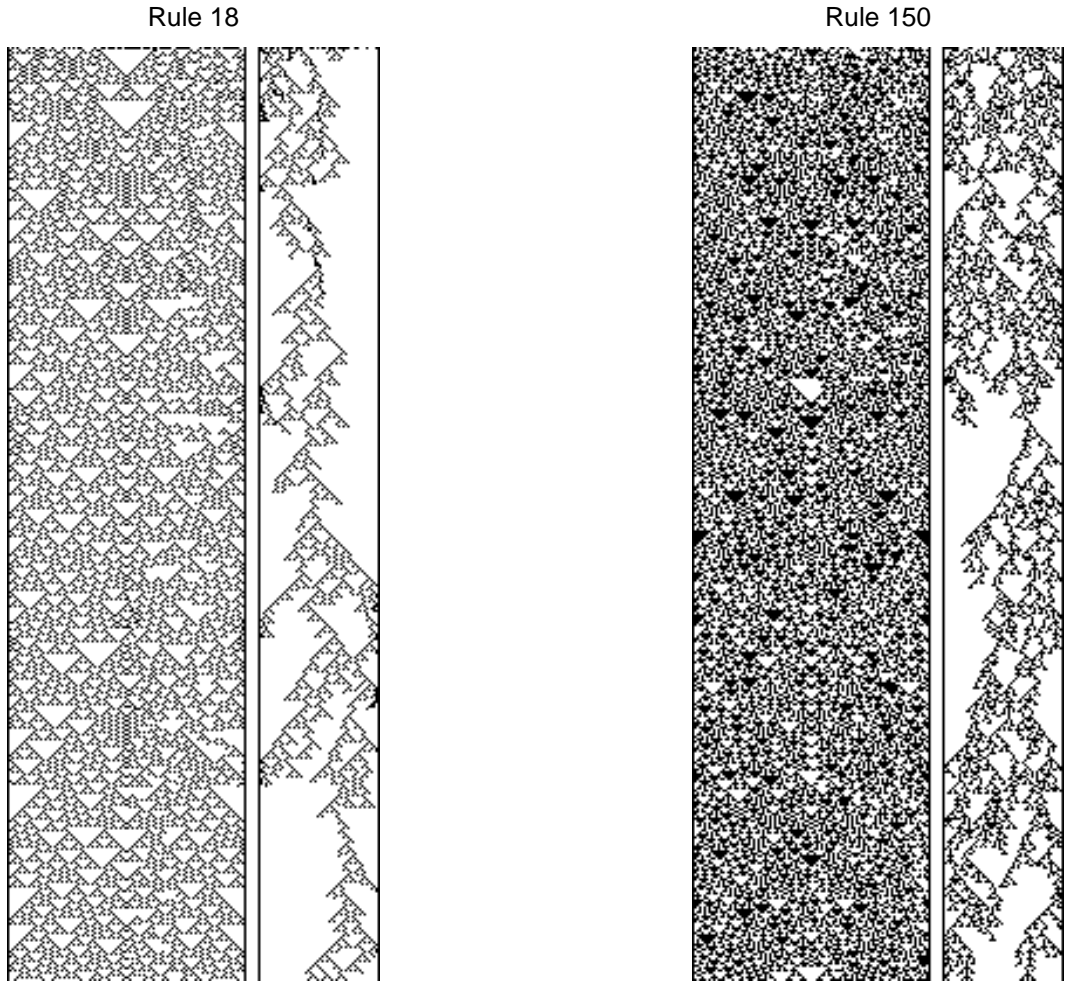}
	\caption{Space-time configurations of automata with $N=100$ sites iterated during 
$T=400$ time steps evolving under rules 18 and 150 for $p \lesssim p_c$. 
Left panels show the automaton evolution in time (increasing from top to bottom) 
and the right panels display the evolution of the corresponding DA.}
\label{fig1}
\end{figure}

\begin{figure}
	\includegraphics{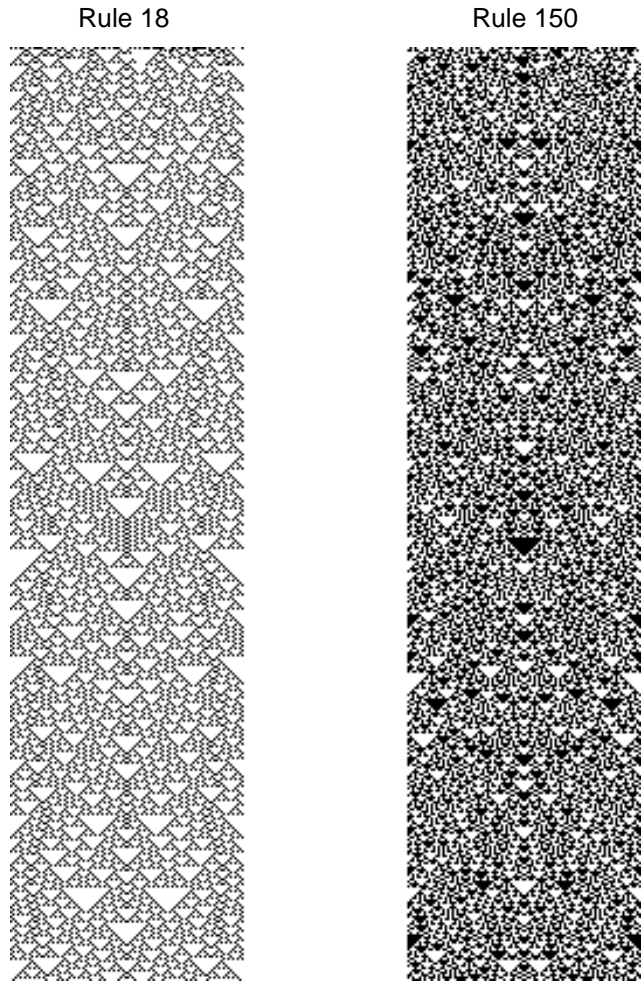}
	\caption{Time configurations of automata with $N=100$ sites iterated during $T=400$ 
time steps evolving under rules 18 and 150 using $p > p_c$. The space symmetry of 
the evolving patterns is clearly visible.}
\label{fig2}
\end{figure}

In Fig. \ref{fig1} the space-time configurations of automata 
evolving under rules 18 and 150 are shown for $p \lesssim p_c$. 
The automata are composed by $N=100$ sites and were iterated during $T=400$ time steps.
Left panels show the automaton evolution in time (increasing from top to bottom) and the 
right panels display the evolution of the corresponding DA. For $p \lesssim p_c$, 
complex structures can be observed in the evolution of the DA. As $p$ approaches 
its critical value $p_c$, the evolution of the DA become more stumped and reminds 
the problem of structures trying to percolate the plane~\cite{Pomeau1986,Sanchez2005}.
In Fig. \ref{fig2} the space-time configurations of the same automata are displayed for $p > p_c$. Now, the
space symmetry of the evolving patterns is clearly visible.

Table \ref{table1} show the numerically obtained values of $p_c$ 
for different rules displaying complex behavior. It can be seen that some rules can not sustain
symmetric patterns unless those patterns are forced to it by fully coupling 
the totality of the symmetric sites ($p_c=1$).  
The rules whose local dynamics verify $\phi(s_1,s_0,s_2)=\phi(s_2,s_0,s_1)$
can evidently sustain symmetric patterns, and these structures are induced 
for $p_c<1$ by the method here explained. 

\begin{figure}
	\includegraphics{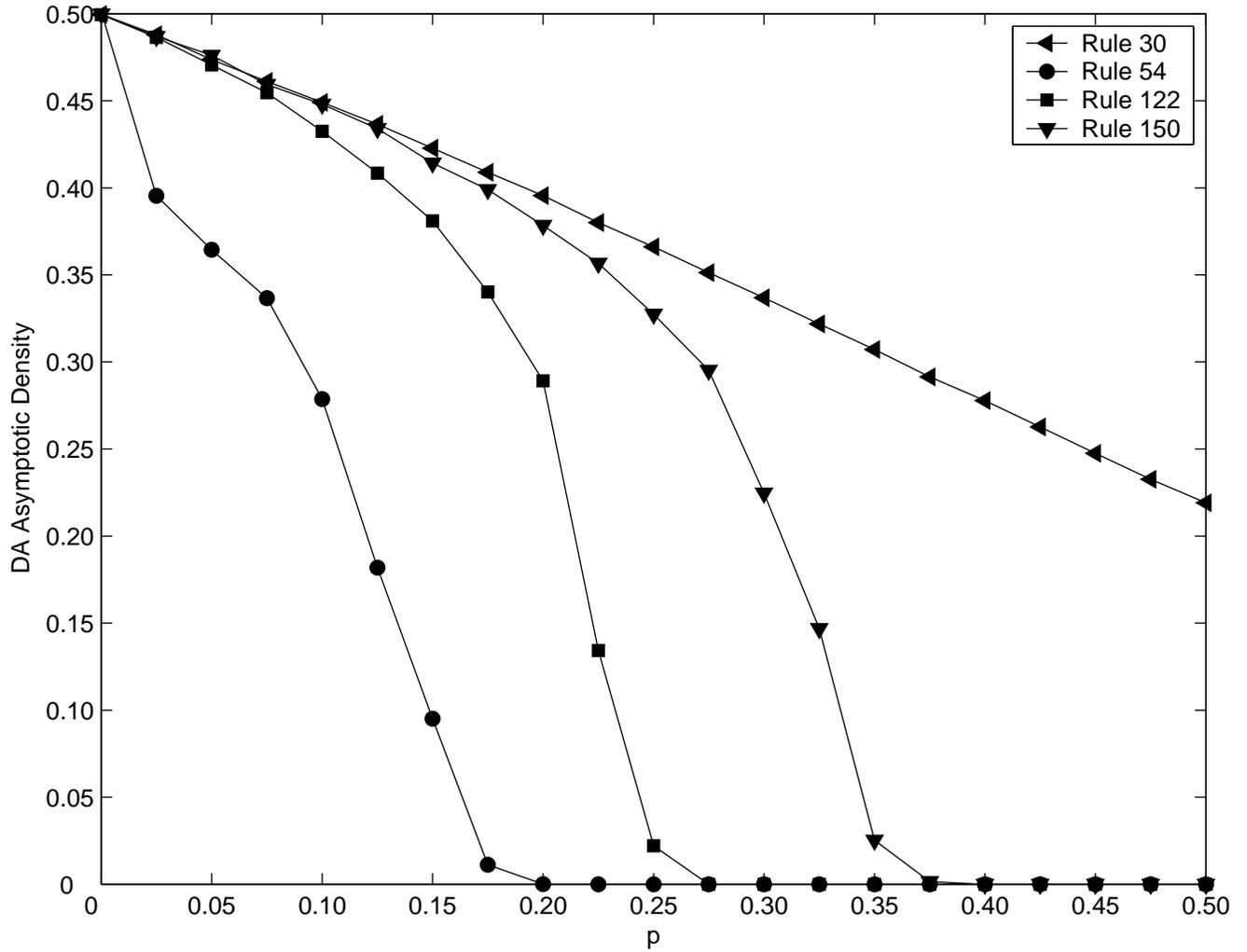}
	\caption{Asymptotic density of the DA for different rules is plotted as a 
function of the coupling probability $p$. Different values of $p_c$ for each 
rule appear clearly at the points where $\rho \to 0$. The automata with $N=4000$ 
sites were iterated during $T=500$ time steps. The mean values of the last 
$100$ steps were used for density calculations.}
\label{fig3}
\end{figure}

Finally, in Fig. \ref{fig3} the asymptotic density of the DA, $\rho^t$ for $t \to \infty$, 
for different rules is plotted as a function of the coupling probability $p$. 
The values of $p_c$ for the different rules appear clearly at the points where $\rho \to 0$.

JRS acknowledges the support of Grant BID-ANPCyT PICT 02-13533 (Argentina) and AUIP (Spain).
RL-R acknowledges the support of Spanish Research Project FIS2004-05073-C04-01.

%%\bibliography{prebr1}

%% ==================================================================================================================
\newpage
\begin{table}%[H] add [H] placement to break table across pages
 \label{table1}
 	\begin{ruledtabular}
 		\begin{tabular}{|c|c|c|c|c|c|c|c|c|c|c|c|c|c|}
 			Rule  & $18$ & $22$ & $30$ & $54$ & $60$ & $90$ & $105$ & $110$ & 
			$122$ & $126$ & $146$ & $150$ & $182$ \\
 			\hline 		
 			$p_c$ & $0.25$ & $0.27$ & $1.00$ & $0.20$ & $1.00$ & $0.25$ & 
			$0.37$ & $1.00$ & $0.27$ & $0.30$ & $0.25$ & $0.37$ & $0.25$
 		\end{tabular}
 	\end{ruledtabular}
 	\caption{Numerically obtained values of the critical probability $p_c$ 
for different rules displaying complex behavior. Rules that can not sustain
symmetric patterns need fully coupling of the symmetric sites, i.e. ($p_c=1$).}
 \end{table}

%% ==================================================================================================================

\newpage
\newpage

\newpage
 
\newpage 

%% ==============================================================================

\end{document}